\documentclass{iopart}

\usepackage{graphicx}

\newcommand{\Co}{CeCoIn$_5$}
\newcommand{\Rh}{CeRhIn$_5$}
\newcommand{\Ir}{CeIrIn$_5$}

\begin{document}

\letter{Field-angle Dependence of the Zero-Energy Density of States in the Unconventional Heavy-Fermion Superconductor {\Co}}

\author{H~Aoki\dag, T~Sakakibara\dag, H~Shishido\ddag, R~Settai\ddag, Y~\={O}nuki\ddag, P~Miranovi\' c\S~and K~Machida\S}

\address{\dag\ Institute for Solid State Physics, University of Tokyo, Kashiwa, 
Chiba 270-8581, Japan}
\address{\ddag\ Graduate School of Science, Osaka University, Toyonaka, 
Osaka 560-0043, Japan}
\address{\S\ Department of Physics, Okayama University, Okayama 700-8530, Japan}


\begin{abstract} Field-angle dependent specific heat measurement has been 
done on the heavy-fermion superconductor {\Co} down to
$\sim0.29~\mathrm{K}$, in a magnetic field rotating in the tetragonal $c$-plane. 
A clear fourfold angular oscillation is observed in the specific heat with the
minima (maxima) occurring along the [100] ([110]) directions.
Oscillation persists down to low fields $H\ll H_{c2}$, thus directly 
proving the existence of gap nodes. The results indicate that the 
superconducting gap symmetry is most probably of $d_{xy}$ type. 
\end{abstract}

\submitto{\JPCM}

\ead{haoki@issp.u-tokyo.ac.jp}

\maketitle

Recently a new class of heavy-fermion superconductors Ce$T$In$_{5}$ 
($T$ = Rh, Ir, and Co) has been discovered. While {\Rh}~\cite{Rh}
is a pressure-induced superconductor, {\Ir}~\cite{Ir} and {\Co}~\cite{1st} 
show superconductivity at an ambient pressure at 0.4 and 2.3~K,
respectively. Among these, unique properties of {\Co} have attracted much 
attentions in these years. Various experiments such as specific 
heat \cite{movshovich01,ikeda}, thermal conductivity~\cite{movshovich01} 
and NMR $T_1$ measurements~\cite{kohori} have revealed that {\Co} is an 
unconventional superconductor with line nodes in the gap. Together
with the suppression of spin susceptibility below $T_{c}$ 
\cite{kohori,curro01}, this compound is identified as a $d$-wave
superconductor. Recent rf penetration depth measurement~\cite{chia03} and
the flux line lattice imaging study by small-angle neutron scattering
\cite{eskildsen03} also seem to be consistent with the existence of line nodes
running along the $c$-axis.

In unconventional superconductors, identification of the gap-node structure is
of fundamental importance in understanding the pairing mechanism. Regarding 
this issue, it has recently been pointed out that the zero-energy density of 
states (ZEDOS) in the superconducting mixed state exhibits a characteristic 
oscillation with respect to the angle between $H$ and the nodal direction 
\cite{Vekhter99,Maki00,Won01}. An intuitive explanation of this effect 
employs a Doppler shift of the quasiparticle (QP) energy spectrum 
due to the local supercurrent flow \cite{Vekhter99,Maki00,Won01,volovik}. 
More quantitative analysis of the effect has been done recently by 
Miranovi\'{c} \etal \cite{cal_amp} within the quasiclassical 
formalism, which incorporates also the contribution of QPs in the core 
states of vortices. Experimentally, the angular dependent ZEDOS can be probed 
by the thermal conductivity~\cite{Yu,izawa,izawa01} or the specific heat
\cite{park} measurements in rotating magnetic field at low $T$. As for {\Co}, 
the angle-resolved thermal conductivity ($\kappa (\theta)$) measurement 
\cite{izawa01} has revealed a clear fourfold oscillation in a magnetic field 
rotating within the $c$-plane. It was argued that the line nodes
exist along the [110] directions ($d_{x^2-y^2}$ type gap symmetry).

The interpretation of the thermal conductivity data is, however, necessarily 
involved because $\kappa (\theta)$ is proportional to the specific heat
$C (\theta)$ as well as to the QP scattering time $\tau (\theta)$.
The latter also depends on the field orientation but with opposite angular
oscillation amplitude to the former.  Which of these two predominantly 
contributes to $\kappa (\theta)$ is a subtle question, making it difficult
to identify the gap node direction in some cases. It is therefore of vital 
importance to directly observe ZEDOS by the angle-resolved specific heat
measurement.  In this Letter, we have examined the $C(H, \theta)$ of {\Co} in 
a magnetic field rotating within the $c$-plane. We observed a clear fourfold 
oscillation in $C(H,\theta)$, and argue that the oscillation originates
from the nodal gap  most probably of $d_{xy}$ symmetry.

A single crystal of {\Co} was grown by the so-called self-flux method
\cite{cry}. The sample was cut into a thin plate ($\sim 2.3\times 2.0 \times 0.7~
\mathrm{mm^3}$) with the largest plane oriented perpendicular to the $c$-axis.
Field-angle-dependent specific heat measurement was done by a standard 
adiabatic heat-pulse technique using a $^{3}$He refrigerator. A transverse 
magnetic field was generated by a  split-pair superconducting magnet and 
applied parallel to the $c$-plane of the sample mounted on a quartz platform. 
The $^{3}$He refrigerator was mounted on a mechanical rotating stage driven 
by a computer controlled stepping motor at the top of the Dewar, by which a 
quasi-continuous change of the field direction with the minimum step of 
0.04$^{\circ}$ could be made. The angular dependent specific heat data were 
collected at an interval of $2^\circ$ in the range $\pm 105^\circ$ with 
respect to the $a$-axis. At the temperature of 380~mK, the 
$2^\circ$-rotation of the refrigerator caused a heating of the sample by 
$\sim100~\mathrm{mK}$ in a field of 5~T. In order to accurately evaluate $C(H,\theta)$ 
at constant temperature, each data point was taken after the sample 
temperature relaxed to within 0.1\% of the initial temperature. The details 
of the experimental set-up will be published elsewhere~\cite{setup}. Field 
dependence of the specific heat up to 12.5~T was also measured by the 
relaxation method using a commercial calorimeter (PPMS, Quantum Design Co.).

\begin{figure}[b]
\begin{center}
\includegraphics[width=8cm]{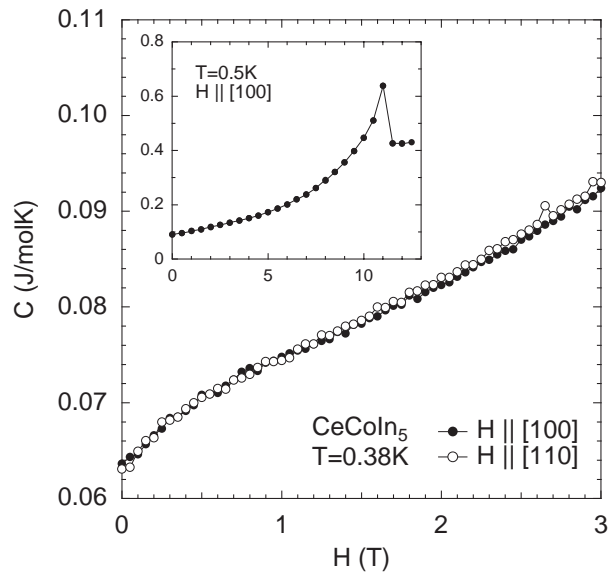}
\end{center}
\caption{\label{CH}Field dependence of the specific heat $C(H)$ of {\Co} for 
$\bi{H} \parallel [100]$ (solid circles) and $[110]$
(open circles), measured at $T=0.38~\mathrm{K}$. Inset: Overall field 
variation of $C(H)$ up to 12.5~T for $\bi{H} \parallel [100]$, 
measured at 0.5~K .}
\end{figure}

\Fref{CH} shows the field dependence of the specific heat $C(H)$ of 
{\Co} obtained at $T=0.38~\mathrm{K}$ for $\bi{H} \parallel [100]$ and $[110]$.
Throughout this paper, the 
nuclear-spin contribution $C_{\rm nuc}$ has been subtracted from
the data, assuming the form $C_{\rm nuc}=(A_0+A_1H^2)/T^2$ with 
$A_0=7.58\times 10^{-2}~\mathrm{mJ\,K\,mol^{-1}}$ and $A_1=6.90\times 
10^{-2}~\mathrm{mJ\,K\,mol^{-1}\,T^{-2}}$~\cite{ikeda}.
$C(H)$ of {\Co} in the superconducting mixed state is quite unusual.
This behavior is very different from the one expected for ordinary 
$s$-wave superconductors in which $C(H)$ linearly increases with $H$ due to
the contribution of QPs trapped in the vortex cores. In many of anisotropic 
superconductors, a power law dependence of $C(H)\propto H^\beta$ with 
$\beta\sim0.5$ has been observed and is attributed to the nodal QP excitations.
The negative curvature of $C(H)$ in low fields seen in \fref{CH} 
is thus consistent with  the existence of gap nodes in {\Co}.  
However, the curvature of $C(H)$ changes sign above 2~T and becomes strongly 
positive at higher fields, as shown in the inset. Similar behavior is 
previously reported for $\bi{H} \parallel [001]$~\cite{ikeda},
as well as in Sr$_2$RuO$_4$ for $\bi{H} \parallel [110]$~\cite{nishizaki}.
Although the reason for this field dependence of the specific heat in {\Co} is unclear at the moment, it might be
related to non-Fermi liquid behavior near the quantum critical point.
With increasing $H$ in the $c$-plane, we observed a small but discernible
anisotropy in $C(H)$ as shown in \fref{CH}; $C(\bi{H} \parallel [100])<C(\bi{H} \parallel [110])$. 

\begin{figure}[t]
\begin{center}
\includegraphics[width=7.7cm]{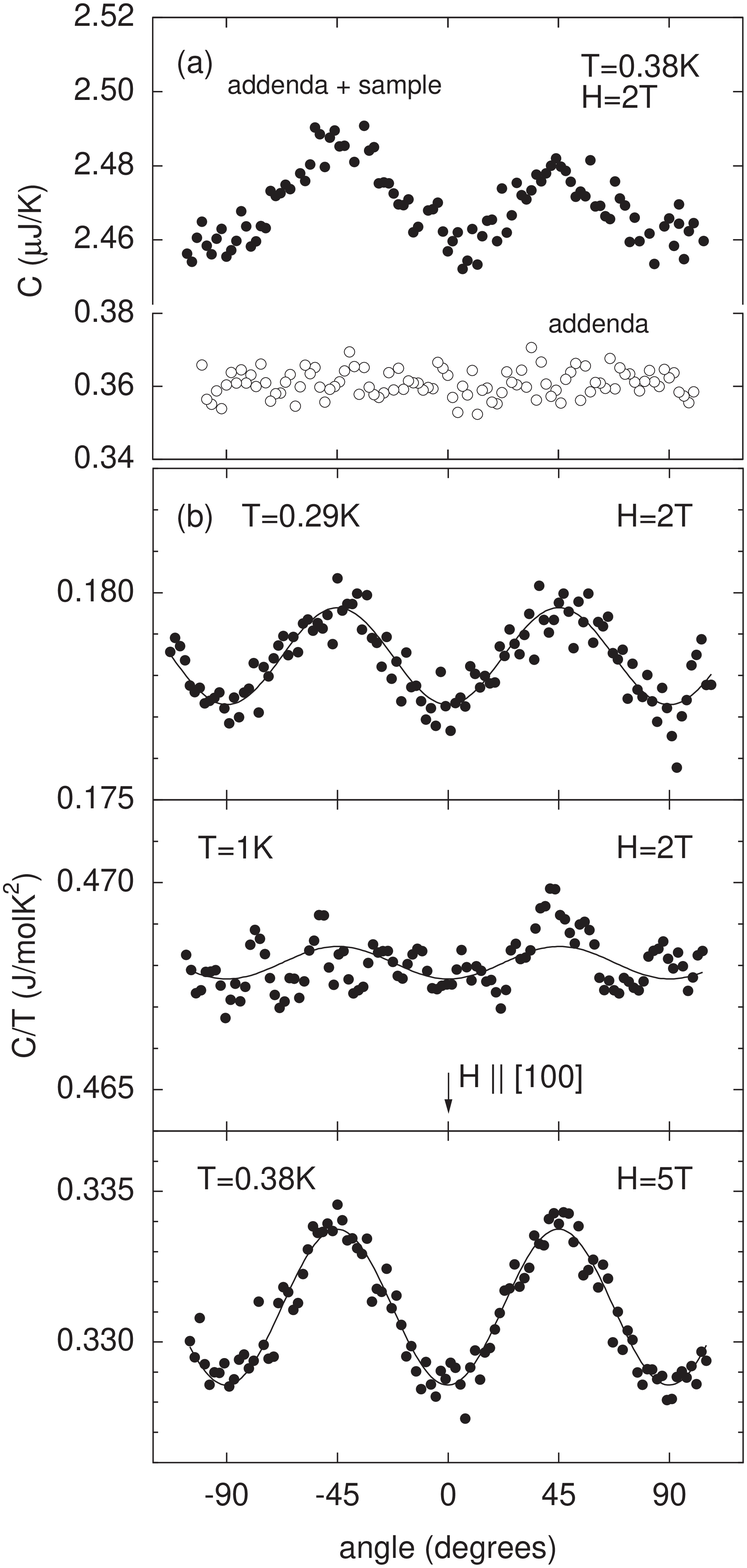}
\end{center}
\caption{\label{signal}(a) Angular dependence of the total specific heat of 
{\Co} (solid circles) mounted on the platform, in a field of 2~T rotated in 
the $c$-plane at $T=0.38~\mathrm{K}$.  $\theta=0$ is the [100] direction. 
Open circles are the results without the sample.
(b) Field-orientational dependence of the specific heat of {\Co} for $H=2~\mathrm{T}$ measured at $T=0.29~\mathrm{K}$ and 1~K, and for $H=5~\mathrm{T}$ measured at $T=0.38~\mathrm{K}$. Solid lines are the fit to $C(H,\theta)/T=(C_{0}+
C_{H}(1+A_{4}\cos 4\theta))/T$,
where $C_{0}=0.0361, 0.0631$ and 0.4042~$\mathrm{J\,mol^{-1}\,K^{-1}}$ for $T=0.29$, 0.38 and 1~K, and
$A_{4}=-0.0217$\,(2~T, 0.29~K), -0.0157\,(5~T, 0.38~K) and -0.0061\,(2~T, 1~K).}
\end{figure}

The in-plane anisotropy of $C(H)$ can be demonstrated more clearly by 
measuring its field-angular dependence, and some of the results are shown in \fref{signal}. First of all, we examined the 
contribution of the addenda (the lower trace of \fref{signal}(a)), which is
composed of the quartz platform, a thermometer (RuO$_2$ chip resistor) and a 
heater. As can be seen in the figure, there is virtually no angular dependence
in the addenda specific heat, implying that the field-orientational dependence
of the thermometer is negligible. By contrast, there is a small but distinct 
fourfold oscillation in the raw specific heat data with the sample on the 
platform (the upper trace of \fref{signal}(a)).  Comparing these two data, it is 
obvious that the observed angular oscillation is intrinsic to the sample. No 
appreciable twofold component is observed in the oscillation, indicating that 
the magnetic field is well oriented along the $c$-plane.

\begin{figure}[b]
\begin{center}
\includegraphics[width=8cm]{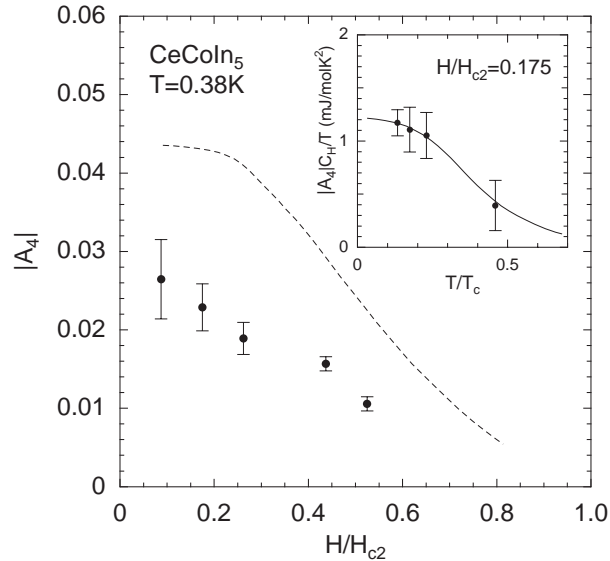}
\end{center}
\caption{\label{amp}Field dependence of the amplitude $|A_{4}|$ of the 
fourfold oscillation at $T=0.38~\mathrm{K}$, plotted as a function of reduced 
field $H/H_{c2}$. The broken line is the calculated field variation for $|A_4|$ 
at $T=0$ assuming the $d_{xy}$-type gap node. Inset: temperature dependence of the absolute amplitude
$|A_{4}|\,C_{H}/T$ measured at $H/H_{c2}=0.175$. The solid line is a guide for the eyes.}
\end{figure}

The specific heat $C(H,\theta)/T$ of {\Co} is shown in \fref{signal}(b) for $H=2~\mathrm{T}$ ($T=0.29$ and 1~K) and 5~T ($T=0.38~\mathrm{K}$) as a function of the field angle $\theta$ measured with respect to the $a$-axis.
The fourfold oscillation is clearly seen at the base temperature of 0.29~K but rapidly fades away at higher temperatures.
$C(H,\theta)$ can be decomposed into a constant and 
field-angle-dependent terms: $C(H,\theta)=C_{0}+C_{H}(1+A(\theta))$. 
$C_{0}$ is the zero-field term mainly due to thermally excited
quasiparticles and phonons, whereas $C_{H}$ and $A(\theta)$ are 
field-dependent. The solid lines in \fref{signal}(b) are the fitting results
assuming a simple form $A(\theta)= A_{4}\cos 4\theta$, by which we evaluated 
the amplitude of the fourfold angular oscillation. The sign of $A_4$ is 
negative, with the minima of $C(H,\theta)$ occurring along the $a$-directions.
In \fref{amp}, we plotted the field dependence of the relative amplitude
$|A_{4}|$ as a function of $H/H_{c2}$.  Most remarkably, $|A_{4}|$ 
decreases monotonously with $H$ within the range of fields examined. The inset
of \fref{amp} shows the temperature variation of the absolute amplitude $|A_{4}|\,C_{H}/T$ measured 
at $H=2$~T ($H/H_{c2}=0.175$). The fourfold angular oscillation in 
$C(H,\theta)/T$ rapidly diminishes with increasing $T$ and vanishes above 
$T_{c}$, implying that superconductivity is responsible for its occurrence.

We now discuss the origin of the fourfold oscillation in $C(H,\theta)$. First 
of all, we should pay attention to the in-plane anisotropy of $H_{c2}$,
which is determined both by the gap topology and Fermi surface anisotropy;
$H_{c2}\parallel [100]$ is about 2.7\% larger than $H_{c2} \parallel [110]$
\cite{cry}. The in-plane anisotropy of $H_{c2}$ alone would also give
rise to the angular dependence of ZEDOS. This is  because the gap amplitude,
which is a decreasing function of the reduced field $H/H_{c2}$, becomes 
field-angle dependent under fixed $H$. This effect certainly becomes 
important when $H$ is near $H_{c2}$, and enhances $C(H,\theta)$ for 
$\bi{H} \parallel [110]$ along which the system is closest to the normal state; 
the sign of the fourfold amplitude is the same as what we observed.
However, in  the limit of low fields $H\ll H_{c2}$ the effect of
the upper critical field can be disregarded.  A crude estimate, assuming 
that $C(H)$ in \fref{CH} can be scaled linearly with $H_{c2}(\theta)$,
shows that the $\sim3$\% anisotropy in $H_{c2}$ would give the fourfold 
term $A_4$ of the order of 1\% at 2~T.  This definitely overestimates the 
effect of $H_{c2}$ anisotropy, but is still smaller than the observed 
one. Moreover, one expects the effect of the upper critical field
to proliferate with increasing field. Instead,  the oscillation
amplitude is a decreasing function of $H$ up to at least 
$H\sim 0.5H_{c2}$.
The strong temperature variation of the fourfold amplitude of $C(H)/T$ shown in the inset of the \fref{amp} is also hardly explained by the $H_{c2}$ in-plane anisotropy effect, because the upturn in $C(H)/T$ does not change much in these temperature regions~\cite{ikeda}.
Since the effect of $H_{c2}$ anisotropy is 
incompatible with the data, we are led to conclude that the observed angular 
oscillation in $C(H,\theta)$ originates from the nodal structure, which has 
fourfold symmetry in the $ab$-plane.

In the mixed state of anisotropic superconductors enhanced contribution to
ZEDOS is coming from the QPs with momentum along the gap node direction.
The enhancement depends on the relative position of the gap and magnetic
field direction. Accordingly, ZEDOS becomes field-angle dependent, and 
exhibits a characteristic oscillation with respect to the angle between $H$ 
and gap-node directions. This ZEDOS oscillation is directly reflected in 
the field-orientational dependence of the specific heat at low $T$, which 
takes minima (maxima) for $H$ parallel to the nodal (antinodal) direction
\cite{Vekhter99,Won01,cal_amp}. The clear minima of $C(H,\theta)$ along [100],
which persist down to low fields, indicate that there are gap nodes 
along these directions. Here we performed the microscopic calculation of 
the field-angle dependent ZEDOS of clean $d$-wave superconductor with 
spherical Fermi surface. We solved numerically and self-consistently the
quasi-classical Eilenberger equations. This formalism is a good approximation 
as long as $k_F\xi\gg 1$ ($k_F$ being the Fermi wave number and $\xi$ 
the coherence length), the condition which is met in {\Co}. It takes into 
account both the Doppler shift effect and the vortex core contribution 
on equal footing and without any further approximation. The result is 
presented with the broken line in \fref{amp}.
Although the predicted curve gives slightly larger amplitude than the
experimental one, it explains the overall field dependence of our 
experimental result reasonably well. The discrepancy between the 
predicted and observed amplitude may partly arise from the fact that the measurements
have been done at finite $T$ ($T/T_{c}\sim0.17$), as inferred from the
strong temperature variation of the amplitude predicted in Ref.~\cite{Won01}. The actual amplitude however gradually levels off below 0.5~K.
Impurities, which are always present, may reduce the 
oscillation amplitude at lower $T$. The present data thus strongly indicate that the gap 
symmetry of {\Co} is most probably of $d_{xy}$ type. 

The present results agree with the $\kappa(\theta)$ measurement on 
{\Co}~\cite{izawa01} in the point that the nodal structure has fourfold 
symmetry in the $c$-plane, but disagrees on the location of the nodes.
$\kappa(\theta)$ shows a fourfold oscillation with the minima along the [010] 
and [100] directions, which is the same oscillation behavior as our 
$C(\theta)$ data. The authors of Ref.~\cite{izawa01} assumed that this 
angular dependence of $\kappa(\theta)$ is dominated by the QP scattering time,
which becomes largest (smallest) when the magnetic field is along the nodal 
(antinodal) directions, just opposite to the ZEDOS contribution. It is 
considered that the QP scattering effect is important at high temperatures, 
but the ZEDOS contribution becomes predominant with decreasing $T$. If the 
angular oscillation of $\kappa(\theta)$ had come from the QP scattering term, 
then its amplitude should change sign on cooling. The observed amplitude of 
the fourfold term in $\kappa(\theta)$ however continues to increase down to 
the lowest $T$ of 0.35~K~\cite{izawa01}. Whether the amplitude changes sign at
still lower temperatures or not would be an interesting issue, but our 
$C(\theta)$ data implies that the ZEDOS effect is already predominant in
$\kappa(\theta)$ at temperatures above 0.35~K.

Present data also does not contradict the neutron scattering experiment
by Eskildsen \etal \cite{eskildsen03}, who find the square
vortex lattice oriented along the [110] direction. The result of
neutron scattering experiment was argued to support the picture
of {\Co} as being the $d_{x^2-y^2}$ superconductor. 
However, the structure and the orientation of the vortex lattice is 
determined by the combined effect of the  Fermi surface anisotropy and the gap 
structure. Which effect will prevail is a subtle question and the answer 
depends on many factors: field value, temperature, degree of Fermi surface 
anisotropy etc. Thus, the orientation of the square vortex lattice in
tetragonal crystals may not serve as a conclusive test for the positions
of gap nodes. Model calculation by Nakai \etal \cite{nakai} shows a 
variety of stable vortex lattice structures in tetragonal $d_{xy}$-wave
superconductors. It is also shown that the low field square vortex lattice is
oriented along [110] direction which is consistent with the neutron 
scattering experiment. It is interesting to point out that the 
square vortex lattice orientation should rotate by $\pi/4$ in high fields
if the $d_{xy}$ is indeed correct identification of the gap function.
  
In summary, we have performed the angle-resolved specific heat measurements 
on the heavy-fermion superconductor {\Co} in a magnetic field rotating in the 
basal $ab$-plane. We observed a clear fourfold symmetry in $C(H,\theta)$ with 
the minima oriented along [100] directions, which comes from the field-angular
oscillation of the zero-energy density of nodal quasiparticles. The results 
imply that the superconducting gap node of {\Co} is located most likely along 
the [100] and [010] directions, suggesting the symmetry to be of $d_{xy}$ 
type. 
\ack
We thank Y~Matsuda, T~Tayama and K~Izawa for stimulating discussions.
We are also grateful to Z~Hiroi for his cooperation in specific heat measurements by PPMS.
This work has partly been supported by the Grant-in-Aid for Scientific Research of 
Japan Society for the Promotion of Science.

\end{document}